\documentclass[]{article}

\usepackage{graphicx}
\usepackage{amsmath}
\usepackage{amssymb}
\usepackage{listings}

\title{Secure and Computationally-Efficient Cryptographic Primitive based on Cellular Automation}
\author{Rade Vuckovac}

\begin{document}

\maketitle

\begin{abstract}
Mageto, a random number generator based on one-dimensional cellular automaton (CA) is presented. Three procedures of secure implementation using Mageto is proposed and discussed. Implementations are very efficient in a wide range of hardware and software scenarios. It includes the advanced application of the Internet of Things (IoT) and cyber-physical systems which are both needed for computationally-efficient cryptographic primitives. Furthermore, the proposed primitive is inherently resistant against the Side Channel Attack (SCA), where many currently available ciphers, such as AES, require additional hardware or software effort to prevent SCA line of attack.
\end{abstract}

\section{Introduction}
Cellular automaton (CA) cryptographic use is limited. One significant reason is performance. A survey of CA stream ciphers~\cite{testa2008investigations} shows that encrypting one megabyte of data requires five seconds at best. One exception is MAG (My Array Generator)~\cite{eSTREAM}, which is a Mageto predecessor. It is a one-dimensional cellular automaton (CA). It belongs to a class of three complexity classification schemes p.12~\cite{ilachinski2001cellular}. That means, nearly every initial state evolves in pseudo-random or chaotic fashion. Two significant attributes which make MAG so exceptional are:
\begin{itemize}
	\item MAG is invariant to the cell size. Both 32-bit and 64-bit cells are investigated, and they show the same behaviour. That fact has a huge impact on performance.
	\item MAG update rule is not entirely boolean and SAT solvers tools, which are generally better than brute force~\cite{testa2008investigations}, could not be applied.
\end{itemize}
Apart from that, there are many more important reasons for revisiting MAG and build Mageto on MAG foundations: 
\begin{itemize}
	\item MAG was the main building block for stream cipher entry (eSTREAM; ECRYPT Stream Cipher Project~\cite{rade_mag}).   That entry did not progress to the second round because of the available analysis at that time. In the second round of decision making, the analysis consisted of two attacks~\cite{mag_QUT, mag_kunzli}. Other attacks~\cite{mag_hd, mag_fisher} are variants of the previous two. More analyses were published~\cite{mirzaeidistinguishing} after round two. Those cryptanalysis contest attacks from~\cite{mag_QUT, mag_hd}. In the same analysis, it suggests that the second type of attacks~\cite{mag_kunzli, mag_fisher} are avoidable using one minor alternation proposed in~\cite{rade_mag_note}. According to the published analysis, at least one MAG variant remains secure.
	\item The second reason for the renewed interest in MAG is excellent software performance, see testing published on eSTREAM web page~\cite{performance_mag}. For a short set of results see Table~\ref{performance}. It compares MAG with some well-known stream cipher algorithms.
	\item Another reason is MAG compactness. While Advanced Encryption Standard (AES) is widely used in symmetrical encryption, the emergence of IoT (The Internet of things), with constrained computation power, limits AES usability in that area. Therefore lightweight symmetrical encryption schemes are sought. MAG hardware footprint is a $512$ bytes of memory for automation state, plus a couple of variables. Operational cost is a cellular automation updating rule consisting of one conditional branching, a couple of exclusive or logical operation, one of one's complement and one addition (five basic operations). That should match an extensive range of IoT hardware limited capabilities. See Listing~\ref{appx}~\ref{magic} for details.
	\item MAG, like other cellular automata, has inherent resistance on side channel attacks (SCA) where AES and many other block ciphers implementations in that regard are relatively complex~\cite{CREAM}. The lightweight cipher designs using S-boxes are affected by SCA as well~\cite{Picek}.
\end{itemize}

Please note that the predecessor of our cryptographic primitive Mageto, namely, MAG, and its design choices for various parameters was never discussed or published before. Thus, in this paper, these issues will be addressed as well.

\begin{table}[h]
	\begin{center}
		\begin{tabular}{|l|c|}
			\hline
			Primitive & Stream (cycles per byte) \\ 
			\hline\hline
			MAG-v3    & 2.20                   \\ 
			TRIVIUM   & 4.14                   \\ 
			Salsa20   & 7.64                   \\ 
			RC4       & 14.52                  \\ 
			AES-CTR   & 18.51                  \\ 
			\hline
		\end{tabular}
	\end{center}
	\caption{An extract of eSTREAM software performance table~\cite{performance_mag}.}
	\label{performance}
\end{table}

The rest of paper is organised as follows: Section~\ref{ca_mag} introduces one dimensional CA and defines Mageto proposition. It also shows where Mageto improves over MAG making the Mageto proposition even simpler. Appendix~\ref{appx} include C language general implementation. Section~\ref{mag_secure} shows three different ways to implement the proposed Mageto securely. Section~\ref{mag_analysis} deals with an analysis of MAG known attacks and how they might impact the Mageto proposal. Section~\ref{disscusion} discusses Mageto and its potential usage to a variety of applications, SCA and input flexibility.

\section{Mageto Cellular Automation}\label{ca_mag}
Mageto is a one-dimensional CA. The concept of CA was first discovered in the 1940s by Stanislaw Ulam and John von Neumann. CA is used as a modelling tool in various scientific fields: computer and complexity science, mathematics, physics and biology. Stephen Wolfram is the first to propose the use of CA (rule 30) in cryptography~\cite{wolfram1984computation}.

\begin{figure}[h!]
	\begin{center}
		\includegraphics[width=1\linewidth]{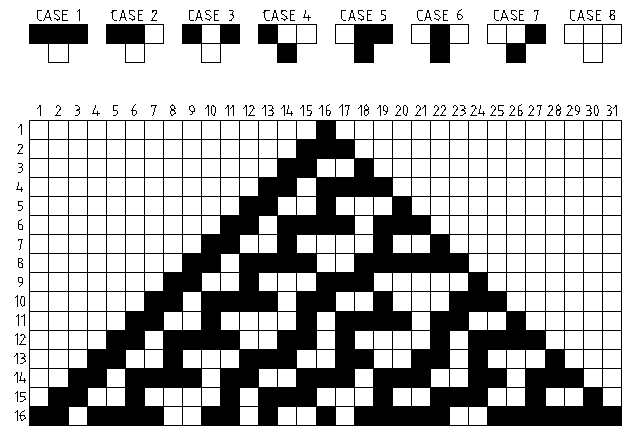}
	\end{center}
	\caption{Rule 30 one-dimensional CA example.}
	\label{rule30general}	
\end{figure}

Figure~\ref{rule30general} showing Wolfram's rule 30 is used to explain the general working of one-dimensional CA. The particular example uses cells with two possible states (black and white). One-dimensional CA initial state is a row. For example, it is the binary string (row 1):
\[
0000000000000001000000000000000
\]
Row 2 is derived from row 1 and so on, with final row 16:
\[
1101111001101001011111001111111
\]
Rules to determine an update of a cell are shown as eight cases. Each case shows one combination of three cells on the top and the derivative cell on the bottom. Three cells from the row above (positioned above right, immediately above and above left) are looked at, and one of the cases are applied to create a cell. For example, a cell from row 2 column 14 is derived by case 8, cell from row 2 column 15 is acquired by case 7, cell from row 2 column 16 is obtained by case 6 and so on. 
Edge cells do not have above left or above right cells to choose a case. In that situation, the first or the last cell from the previous column are used for the ruling. For example, left edge cell from row 16 column 1 looks at cells from row 15, columns 31, 1, 2 and use case 7 to create the cell. The right edge cell from row 16 column 31 use the cells from row 15 columns 30, 31, 1 for lookup and case 4 for cell determination.
Derived rows (rows $2, 3, ...$) are outputs of CA.

If a random stream of bits is required, column $16$ could be used: 
\[
1101110011000101...
\]
A stream generated in this fashion passes many randomness statistical tests and it is used as a random number generator in Wolfram's Mathematica software.

Mageto is also a one-dimensional CA, but it differs from the example by:
\begin{itemize}
	\item Cells are multi-bit words (32 or 64 bits in size).
	\item The update of Mageto cells is serial, left to right, because the rule needs the outcome of the previous cell update (carry).
	\item The Mageto rule also appears to be invariant to cell size because it shows the same random behaviour for $32$ and $64$ bit size cells.
\end{itemize}

Mageto CA is governed by the CA rule and the state of neighbouring cells. A row of cells, in a Mageto case array of elements, are updated from left to right. One evolution cycle is when all items in the array are updated. The next generation is another evolution cycle and so on. The original and modified parameters are shown in Table~\ref{magparam}.

\begin{table}[h]
	\begin{center}
		\begin{tabular}{|l|c|}
			\hline
			MAG		& Mageto \\ 
			\hline\hline
			$a=127$ 				& $a=128$ 					\\ 
			$b=32$  				& $b=32$ 					\\ 
			$c$ calculated			& $c=987654321$ 			\\ 
			$d=0x11111111$ (HEX) 	& $d=010101...$ (binary)	\\ 
			$e=1.5a$ 				&  $e=4a$ 					\\
			\hline
		\end{tabular}
	\end{center}
	\caption{MAG / Mageto parameters.}
	\label{magparam}
\end{table}

Mageto parameters definitions:
\begin{itemize}
	\item The number of cells is $a=128$. That choice forces $2^{128}$ possible execution paths during one evolution cycle and assumes at least $128$ bit level security if the stream is used as a basis for the cipher. 
	\item The cell size is $b=32 \, bits$. The Mageto CA rule appears to be invariant concerning the cell size. $b=64 \, bits$ is used and tested for randomness~\cite{vuckovac2013new} and there are no biases, although the performance doubles because the same generating cost produces double the stream.
	\item Instead of taking the carry value from the array element in the original version, it is initialised as $c=987654321$ (decimal). The value of $987654321$ was chosen, and there is no special meaning behind this choice.
	\item The constant $d$ is now initialised as $d=01010101...$ ($32 \, bits binary$). Again there is nothing special in the constant value. In MAG $d$ was described as an arbitrary value (which it is). Related to this, an initialisation attack~\cite{mag_QUT, mag_hd} where $d=0$ was proposed. It eliminates any adding procedure which simplifies the whole process significantly. The analysis~\cite{mirzaeidistinguishing} showed that any non zero value of $d$ is sufficient to prevent initialisation attacks.
	\item In a modified version, the mixing period $e$ is four evolutions $e=4*a=512$. It assures proper mixing because the original one and half evolutions occasionally produced biases in the first couple of generated rows. The same could be observed in the rule 30 case (Figure~\ref{rule30general}) where the first several rows still retain some patterns.     
	\item The seed $f$ is any binary string equal or smaller than a row of cells $f_{size}\leq a*b$ and $f=k+s+...$ meaning that the key $k$, the salt $s$ and $...$ (IV, pepper and so on) are concatenated to form seed $f$.   
\end{itemize}

\begin{figure}[h]
	\begin{center}
		\includegraphics[width=0.7\linewidth]{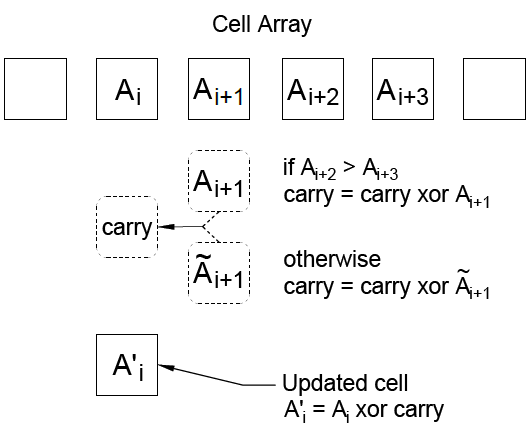}
	\end{center}
	\caption{Mageto cell update rule.}
	\label{mag_rule}	
\end{figure}

Mageto operation is divided into initialisation and update.\\
Mageto initialisation; Originally, the array of $128$ elements $32$ bits wide is initialised to $0$. The seed $f$ is repeatedly concatenated until the resulting concatenation is equal or greater in size than the array. MAG array's first $127$ elements are the initial row, and remaining elements becomes the carry $c$. For example, the seed is $f=seed$ and array is $10$ cells (one byte each), the resulting initial array will be: 
\[
seedseedse
\]
In the Mageto modified version, the seed $f$ is simply copied to zero initialised array and the carry is given as an initial value $c=987654321$. The pattern, with $f=seed$ and array of $10$ bytes, looks like: 
\[
seed000000
\]
Mageto update; when initialised, the rows are created by updating cells from left to right. The edge cases ($A_i$ happens to be on the array end) are handled as rule 30 edge cell cases.  The rule elements are shown in Figure~\ref{mag_rule}. The cell update consists of three steps:
\begin{itemize}
	\item 1st step is to create a new state of carry $c'$.
	\begin{equation}
	c'= 
	\begin{cases}
	c \oplus A_{i+1},& \text{if } A_{i+2}>A_{i+3}\\
	c \oplus \overline A_{i+1},& \text{otherwise}
	\end{cases}
	\end{equation}
	Carry $c'$ is updated by $\oplus$ (exclusive or) with previous value of $c$ and one state of the first element to the right $A_{i+1}$, depending on the relation between the other two cells on the right ($A_{i+2},A_{i+3}$). States of $A_{i+1}$ are: current value ($A_{i+1}$) or it's one complement ($\overline A_{i+1}$). In one evolution cycle (whole array is updated) each cell is changed once and carry is calculated for every cell transformation. Note that the first cell update uses the initial value of carry $c=987654321$.
	\item 2nd step is the actual change of element $A_i$ to $A'_i$.
	\begin{equation}
	A'_{i}=A_{i} \oplus c' 
	\end{equation}
	\item 3rd step is updating current $c'$ value for next cell transformation.
	\begin{equation}
	c'= c'+ d
	\end{equation}
\end{itemize}

\section{Mageto as A Secure Stream Generator}\label{mag_secure}    
Knowledge of the array state renders the Mageto algorithm cryptographically unsound. Three strategies for concealing CA state to make a secure stream from Mageto, are proposed and discussed.

\subsection{Reducing output (Mageto-v1)} \label{reducing}
One of the concealing methods was already used by S. Wolfram on his cellular automata rule 30 (section 10.10\cite{Wolfram2002}). For example, column 16 from Figure~\ref{rule30general}
\[
1101110011000101...
\]
is transformed by taking bits $1,\,3,\,5,\,7,\,9,\,...$ and producing secure stream
\[
1\;0\;1\;0\;1\;0\;0\;0\,...
\]
The similar approach was used in the MAG eSTREAM proposal. The small change in operation is added to achieve a secure stream property. Instead of copying every updated cell to the stream, only the first byte of the cell is added to the secure stream $s$ (now array of bytes).

\begin{figure}[h]
	\begin{center}
		\includegraphics[width=0.7\linewidth]{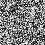}
	\end{center}
	\caption{Mageto evolution history; one pixel represents one byte (256 grey scale).}
	\label{random}	
\end{figure}

For example, Figure~\ref{random} shows updated cells as a stream of bytes where each pixel represents a byte and four bytes are an updated cell. To make a secure stream every 5th byte (every 5th pixel from left to right) is taken and fed to the secure stream. That is bytes:
\[
1,\,5,\,9,\,13,\,17,\,...
\]
Consequently $b_1b_5b_9...$ bytes create a secure stream. This approach was found insecure, and details are shown in Section~\ref{mag_analysis}.\\
Again the small change in pattern avoids weaknesses from the previous design. If the series of bytes taken is changed to: from first cell first byte, second cell second byte, third cell third byte, fourth cell fourth byte, fifth cell first byte and so on... 
\[
1,\,6,\,11,\,16,\,17,\,22,\,27,\,32...
\] 
The stream of bytes $b_1b_6b_{11}b_{16}b_{17}b_{22}b_{27}b_{32}...$ from Figure~\ref{random} will form a secure stream.

\subsection{Combining streams (Mageto-v2)}\label{combo}
One way of making a stream secure is to combine two or more streams. For example, LFSR (Linear-feedback shift register) outputs were combined to make a shrinking generator (planned to be used as a stream cipher~\cite{coppersmith1993shrinking}). The shrinking generator uses two streams, one is a source, and the other is used to decide which bits of the source stream is output.

In the Mageto case the idea is to apply exclusive or between two generated streams (stream $\alpha$ and $\beta$) to produce a secure stream $s$:

\begin{equation}
s_i=\alpha_i\oplus \beta_i
\end{equation}

\begin{table}[h]
	\begin{center}
		\begin{tabular}{|l|c|}
			\hline
			Mageto stream &	stream IV \\ 
			\hline\hline
			$\alpha$ stream  & 1234567890/3 \\ 
			$\beta$ stream  & 9876543210/3  \\ 
			$\gamma$ stream    & ...           \\
			\hline
		\end{tabular}
	\end{center}
	\caption{Combined stream parameters.}
	\label{combined}
\end{table}

One implementation attempt at combined MAG approach was published here~\cite{vuckovac2013new}. It includes source code. The easiest way to implement combined Mageto is to initialise streams separately. Table~\ref{combined} contains initial parameters.
The seed $f$ for each stream now includes corresponding IV. For example, the seed for $\alpha$ is $f_\alpha=k+s+IV$ where $k$ is the key, $s$ is the salt, $IV$ initialisation vector from Table~\ref{combined} and $+$ is the concatenation of strings.

\subsection{Stream Masking (Mageto-v3)}\label{mask_stream}
The idea with masking is to combine (xor) Mageto output with a secret string. The original idea is to use a key as the secret string but Dr Daniel J. Bernstein and prof.dr.Tanja Lange noted that the same attacks (\cite{mag_kunzli, mag_fisher}) apply for that proposal as well.
Alternatively, the secret could be sourced from the execution path history. The branching from the previous evolution can generate string $m$. In the case of $128$ elements array, there is $128$ branching in one evolution cycle making a $128$ bit string $m$. Bits of $m$ are determined by branching; \textit{if} branch will concatenate $0$ and \textit{else} $1$ to the mask $m$. From $m$ the four $32$ bit element mask array $M$ is created.
 
\begin{equation}
M=[B_1; B_2;B_3;B_4]
\end{equation}

The secure stream $s$ is now obtained by exclusive or result from Mageto output cells ($A$) and mask array of four elements ($M$):
\begin{equation}
s_i=A_{i\%a} \oplus B_{i\%4}
\end{equation}
The new mask $M$ is calculated for every evolution cycle and is used as a mask for next cycle. 

\section{Mageto Analysis}\label{mag_analysis}
Mageto-v1; The eSTREAM proposal to strengthen CA MAG was to output just the first byte from the every updated cell. It did not work.  CA can continue updating on just that part of the cell without information from the hidden part:
\begin{itemize}
	\item Since the first bytes of $A_{i\%a}$ and $A'_{i\%a}$ are known the first byte of $c'$ is known as well because:
	\begin{equation}
	A'_{i\%a}=A_{i\%a} \oplus c'
	\end{equation}
	\item Knowing the first byte of $c$, predicting the next unknown value of the carry $c'$ takes guessing the branching outcome. That guess is even easier because of knowledge of the first byte which is compared. 
\end{itemize}

\begin{table}[h!]
	\begin{center}
		\begin{tabular}{|l|c|}
			\hline
			Series of bytes used in the stream & Status \\ 
			\hline\hline
			$1,\,5,\,9,\,13,\,17,\,...$  &   $\times$ \\ 
			$1,\,2,\,7,\,8,\,9,\,10,\,15,\,16...$  &   $\times$ \\ 
			$1,\,6,\,11,\,16,\,17,\,22,\,27,\,32...$  &\checkmark\\ 
			\hline
		\end{tabular}
	\end{center}
	\caption{Mageto secure stream extraction patterns.}
	\label{amenl}
\end{table} 

This kind of attack is detailed in~\cite{mag_kunzli, mag_fisher}.
Two amendments are proposed in~\cite{rade_mag_note} to avoid this line of attack. The idea was to alternate extraction points because the original design did not hide the evolution of the first-byte cell.
Table~\ref{amenl} shows various extraction patterns for the proposed MAG secure stream. The first amendment (second-row Table~\ref{amenl}) is broken as well. The one gap between exposed bytes of the cell did not prevent the same attack although the guessing cost was increased~\cite{mirzaeidistinguishing}. The second amendment (third-row Figure~\ref{amenl}) does have three gaps between visible bytes and is still resisting analysis.

Mageto-v2; The result from exclusive or of two Mageto streams is secure stream $s$. The relevant relations of knowns and unknowns are shown below. Stream $s$ is known and streams $\alpha$ and $\beta$ and carry $c$ are unknown.
\begin{itemize}
	\item [] $\text{row wise  }\,s_i=\alpha_i\oplus \beta_i;\,\, s_{i+1}=\alpha_{i+1}\oplus \beta_{i+1};\, ...$
	\item [] $\text{evolution wise  }\, c'_{\alpha i} \oplus c'_{\beta i}=s_i \oplus s'_{i}$
\end{itemize}
			
Both relations are used to attack Mageto Subsection~\ref{reducing}~\cite{mag_kunzli,mag_fisher}.
Additional streams could be included to strengthen two stream variant. The combination of three Mageto streams producing secure stream might look like:
\begin{equation}
s_i=\alpha_i \oplus \beta_i \oplus \gamma_i
\end{equation}

Mageto-v3; The relations with known $s$ only are shown below concerning the Mageto attacks~\cite{mag_kunzli, mag_fisher}.
\begin{itemize}
	\item [] $ \text{row wise }\, s_i \oplus s_{i+4}=A_i \oplus B_i \oplus A_{i+4} \oplus B_i = A_i \oplus A_{i+4} $
	\item [] $ \text{evolution wise  }\, c'_i= s_i \oplus s'_i=A_i \oplus B_i \oplus A'_i \oplus B'_i  $
\end{itemize}

\section{Mageto Advantages} \label{disscusion}

In this section a few important Mageto features are discussed. One highlight is applicability with the respect to various hardware platforms. There is also Mageto resistance to the side channel attack and Mageto flexibility to the inputs other than key.  

\subsection{Mageto Implementation and Performance}
Three Mageto cipher variants are presented:
\begin{itemize}
	\item Mageto-v1 from subsection~\ref{reducing} is the simplest. For producing one byte, only the set of operations from Table~\ref{steps} are needed plus the overhead of extracting byte from the cell $A'$ and array navigation.
	\begin{table}[h]
		\begin{center}
			\begin{tabular}{|l|}
				\hline
				Mageto single cell update\\ 
				\hline\hline
				$
				c'= 
				\begin{cases}
				c \oplus A_{(i+1)},& \text{if } A_{(i+2)}>A_{(i+3)}\\
				c \oplus \overline{A}_{(i+1)},& \text{otherwise}
				\end{cases}
				$\\
				$A'_{i}=A_{i} \oplus c'$\\
				$c'=c'+d$\\ 
				\hline
			\end{tabular}
		\end{center}
		\caption{The operations needed for one step cell update.}
		\label{steps}
	\end{table} 
	The array containing Mageto cells is only $512$ bytes in size and with the mentioned set of operation, lowers the entry hardware requirements bar significantly. That includes a wide array of IoT implementations. The Mageto-v1 performance, the same efficiency as mag-v1, is very comparable with AES, see Table~\ref{performance1}.   
	\begin{table}[h]
		\begin{center}
			\begin{tabular}{|l|c|}
				\hline
				Crypto primitive & Stream (cycles per byte) \\ 
				\hline\hline
				RC4       & 14.52                  \\ 
				AES-CTR   & 18.51                  \\ 
				MAG-v1    & 20.43                   \\ 
				\hline
			\end{tabular}
		\end{center}
		\caption{An extract of eSTREAM software performance table~\cite{performance_mag}.}
		\label{performance1}
	\end{table}
	\item Mageto-v2 approach subsection~\ref{combo}, is a little bit more complex. On the other hand, efficiency is improved. By roughly doubling effort, output increases four times. That is $4$ bytes per $2$ CA steps comparing $1$ byte per $1$ CA step (subsection~\ref{reducing}). There are other ways to improve performance. One way relies on the fact that combining streams could be created in parallel. By that technique performance is $4$ bytes / $1$ CA step ($2$ steps in parallel). This technique also enables adding streams if needed without affecting performance. Another efficiency approach could be increasing the size of the CA cell from $32$ to $64$ bits producing $8$ bytes per step. For details see Table~\ref{perf_combo}.
	\begin{table}[h]
		\begin{center}
			\begin{tabular}{|l|c|}
				\hline
				MAG variants			   & Performance		   \\ 
				\hline\hline
				Mageto-v1                 & $1byte/1step$ \\ 
				Mageto-v2                 & $2byte/1step$ \\ 
				Mageto-v2 parallel        & $4byte/1step$ \\ 
				Mageto-v2 parallel 64 bit & $8byte/1step$ \\ 
				Mageto-v3 		           & $4byte/1step$ \\ 
				Mageto-v3 64 bit 		   & $8byte/1step$ \\
				\hline
			\end{tabular}
		\end{center}
		\caption{Various performance details.}
		\label{perf_combo}
	\end{table}
	\item Mageto-v3 from subsection~\ref{mask_stream} is a notch more complex than previous variants. Developing mask for each evolution step is the reason. This approach also improves performance concerning Mageto-v1. That improvement does not need parallelism. Although Mageto-v3 is relatively more complex, it is still significantly simpler than AES from a hardware and software point of view, delivering better performance. Table~\ref{performance} shows mag-v3 (the same as Mageto-v3) 64-bit implementation versus AES and other primitives. Table~\ref{perf_combo} shows performances between various Mageto variants where one step from Table~\ref{steps} produces $1-8\, bytes$ towards secure stream depending on the variant used.
\end{itemize}

\subsection{Side Channel Resistance}
When discussing side channel attack on AES, the quite often cited work~\cite{bernstein2005cache} cannot be avoided. There is an assertion mentioned in the abstract that attacks come from the AES design flaw rather than AES implementation. Furthermore, this report is also a call for research into functions with constant time execution. Mageto as CA appears to have the rule which runs in constant time, but some attention is still needed. When implementing Mageto in SCA resistant mode, the algorithm branching structure should be addressed. There are two issues:
\begin{itemize}
	\item The first one is to compare secret cells in constant time. By measuring comparing operations, some properties of the Mageto cells could be determined. For example, comparing the equal cells byte by byte will take the longest time to execute. The solution is to use constant time comparison functions. Some cases can be found here~\cite{CR}. Note, the solutions are not entirely portable therefore finished program assemblies for particular hardware should be checked for correctness in any case.
	\item The second Mageto issue is branching execution times. For example, Mageto timing for each branch differs, consequently the mask parity of Mageto-v3~\ref{mask_stream} could be readily determined. The solution is equal execution time for each branch. That is accomplished by introducing two pre-calculated intermediate variables $v$ and $w$ before branching. The update of Mageto carry with $v$, and $w$ has equal execution of branches:
	
	\begin{equation}
	v=A_{i+1}
	\end{equation}
	
	\begin{equation}
	w=\overline{A}_{i+1}
	\end{equation}
	
	\begin{equation}
	c'= 
	\begin{cases}
	c \oplus v,& \text{if } A_{i+2}>A_{i+3}\\
	c \oplus w,& \text{otherwise}
	\end{cases}
	\end{equation}
\end{itemize}

\subsection{Input Flexibility}

Generally, there is a requirement of an initialisation vector (IV) in symmetrical encryption. For example, AES CBC (AES in Cipher Block Chaining) needs a unique $128\text{ bit}$ IV for every message processing. If more than $128\text{ bit}$ IV is required for some reason, the key derivation function (KDF) is necessary to deliver properly sized key and IV.

In that respect, Mageto allows an additional $480\text{ bytes}$ for a nonce, salt, pepper and so on... if needed. It can be used for resisting various repeat attacks for example. That can be accomplished without using KDF as it is the case with AES.

\subsection{Conclusion}

Mageto offers an entirely new cryptographic primitive. It has a straightforward and compact implementation. It also provides SCA resistance which is very important for not physically secured hardware such as IoT. Variant Mageto-v1 also benefits from the fact that its predecessor mag-v1 had in-depth analysis and remains secure~\cite{mirzaeidistinguishing}.

\bibliographystyle{plain}
\bibliography{mageto} 

\appendix
\section{Mageto inplementation details}\label{appx}

Implementation is written in C language and source code is in Listing~\ref{magic}. 
\begin{lstlisting}[label=magic, language=C]
#include <stdio.h>
#include <stdint.h>
#include <string.h>
#include <stdlib.h>

void stir(uint32_t * array, uint32_t updates, int init)
{
	const uint32_t mixer = 1431655765;			//010101...
	uint32_t carry = 987654321;
	int a = 128;								//array size
	uint32_t i;			
	for(i = 0; i < updates; i++)
	{
		if(array[(i+2)%a]>array[(i+3)%a])
			carry ^= array[(i+1)%a];
		else
			carry ^= ~array[(i+1)%a];

		array[i%a] ^= carry;
		carry += mixer;

		if(init == 1)
			fwrite(&array[i%a] , sizeof (array[i]) ,1 , stdout );
	}
}

int main(int argc, char **argv)
{
	uint32_t bytesfour;
	uint32_t skip = 512;
	uint32_t array[128] = {0};
	
	size_t length = strlen (argv[2]) ;

	bytesfour = atoi(argv[1]);
	memcpy(array, argv[2], length);

	stir(array, skip, 0);
	stir(array, bytesfour, 1);
		
	return 0;
}

\end{lstlisting}

Lets assume that the Listing~\ref{magic} is compiled to $mageto$. The execution command might look like Figure~\ref{fig:command}:

\begin{figure}[h]
	\centering
	\includegraphics[width=0.7\linewidth]{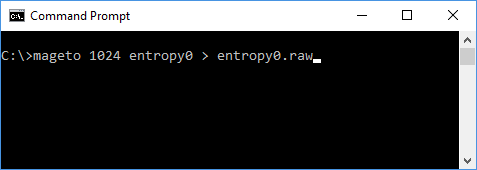}
	\caption{How to run mageto.}
	\label{fig:command}
\end{figure}

Where the first argument $1024$ represents how many cells are sent to the standard output $(4096\, bytes)$. The second argument is the seed $entropy0$ and $>\, entropy0.raw$ redirects standard output to the file.

\begin{figure}[h!] 
	\centering
	\includegraphics[scale=3]{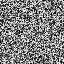}
	\caption{Seed "entropy0", (argv[2]).}
	\label{random_slika}
\end{figure}

\begin{figure}[h!] 
	\centering
	\includegraphics[scale=3]{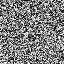}
	\caption{Seed "entropy1", (argv[2]).}
	\label{random_slika1}
\end{figure}

Figure~\ref{random_slika} and~\ref{random_slika1} are showing graphical representations of the outputs when input differs by one bit only. The seeds are strings "entropy0" and "entropy1" respectively. Every pixel is one byte shown in 8-bit grayscale.	
	
\end{document}